# Risk of re-identification for shared clinical speech recordings

Daniela A. Wiepert[1], Bradley A. Malin[2,3,4], Joseph R. Duffy[1], Rene L. Utianski[1], John L. Stricker[1], David T. Jones[1], and Hugo Botha[1]

[1]*Department of Neurology, Mayo Clinic (Rochester), Rochester, MN, 55905, USA*
[2]*Department of Biomedical Informatics, Vanderbilt University Medical Center, Nashville, TN, 37203, USA*
[3]*Department of Biostatistics, Vanderbilt University Medical Center, Nashville, TN, 37203, USA*
[4]*Department of Computer Science, Vanderbilt University Medical Center, Nashville, TN, 37232, USA*

## Abstract

Large, curated datasets are required to leverage speech-based tools in healthcare. These are costly to produce, resulting in increased interest in data sharing. As speech can potentially identify speakers (i.e., voiceprints), sharing recordings raises privacy concerns. We examine the re-identification risk for speech recordings, without reference to demographic or metadata, using a state-of-the-art speaker identification model. We demonstrate that the risk is inversely related to the number of comparisons an adversary must consider, i.e., the 'search space'. Risk is high for a small search space but drops as the search space grows (*precision* > 0.85 for < $1*10^6$ comparisons, *precision* < 0.5 for > $3*10^6$ comparisons). Next, we show that the nature of a speech recording influences re-identification risk, with non-connected speech (e.g., vowel prolongation) being harder to identify. Our findings suggest that speaker identification models can be used to re-identify participants in specific circumstances, but in practice, the re-identification risk appears small.

## Introduction

Advances in machine learning and acoustic signal processing, along with widely available analysis software and computational resources, have resulted in an increase in voice and speech (hereafter 'speech' for simplicity) based diagnostic and prognostic tools in healthcare [1]. Applications of such technology range from early detection of cardiovascular [2], respiratory [3], and neurologic [4] diseases to prediction of disease severity [5] and evaluation of response to treatment [6]. These advances harbor substantial potential to enhance patient care within neurology given the global burden of neurologic diseases [7, 8], poor global access to neurologic expertise [9, 10], and the established role of speech examination within the fields of neurology and speech-language pathology [11].

Large, curated datasets are needed to harness the advances in this area. However, these datasets are costly to assemble and require rare domain expertise to annotate, leading to increased interest in data sharing among investigators and industry partners. However, given the potential identifiable nature of voice or speech recordings and the health information contained within such recordings, significant privacy concerns emerge. For many datasets, conventional de-identification approaches which remove identifying metadata (e.g., participant demographics, date and location of recording, etc.) are sufficient, but sharing speech recordings comes with additional risk as the speech signal itself has the potential to act as a personal identifier [12, 13, 14]. In recognition of this potential problem, voiceprints are specifically mentioned as an example of biometric identifiers with respect to the HIPAA Privacy Rule [15, 16]. Approaches that involve modifying non-linguistic aspects of speech through distortion or alteration of the signal may address the inherent identifiability of the speech signal (i.e., its potential as a voiceprint) [13, 17], but this is not an option when a central part of speech examination in medicine is to use the acoustic signal to detect subtle non-linguistic abnormalities indicative of the presence of neurologic disease [13, 11]. De-identification in compliance with HIPAA may still be possible under the Expert Determination implementation, whereby the risk of re-identification for the unmodified speech recordings is deemed low according to accepted statistical and/or scientific principles [16, 15]. In this respect, various prior studies have investigated the risk of re-identification in research cohort datasets based on demographic or other metadata that may link a participant to their corresponding records [18, 19, 20], but none have explicitly assessed the inherent risk of the acoustic signal itself. Determining the risk of re-identification for recordings in speech datasets and learning how to best mitigate such risk is necessary for healthcare institutions to protect patients, research participants, and the institutions themselves.

Unfortunately, the same machine learning advances that facilitate the use of speech in healthcare have also made adversarial attacks, such as de-anonymization or re-identification attacks, more feasible. For example, attempting to re-identify a speaker from only a speech recording relies on the mature, well-researched field of speaker identification [21, 22]. Studies using speaker identification suggest the potential for identification from the acoustic signal alone is high [23], though there have been minimal studies in the context of adversarial attacks that may result in potential harm to a speaker [24, 25]. Only one prior study has relied upon a speaker identification model for re-identification, and the results suggested that the risk was high with a single unknown, or unidentified, speaker and a moderately small reference set of 250 known, or identified, speakers [25]. As such, the risk inherent in the acoustic signal, devoid of metadata, is non-zero but relatively unknown, and the feasibility for larger datasets is unexplored.

In addition, these approaches are rarely applied to medical speech datasets [26]. This presents a gap in research as medical speech recordings differ from normal speech recordings in a few systematic ways. First, the recordings typically contain abnormal speech (i.e., speech disorders) which may make re-identification harder since many speech disorders are the result of progressive neurologic disease, which cause changes in speech that evolve over months to years [11]. Matching recordings from a time when a speaker was healthy or mildly affected to recordings where they have a more severe speech disorder may be more difficult [27, 28, 29]. Second, the premise of speaker identification is that there are recognizable between-speaker differences tied to identity. However, in a cohort enriched with abnormal speech, a substantial proportion of the variance would be tied to the underlying speech disorder as these cause recognizable deviations [11] resulting in speakers sounding less distinct [30]. Finally, medical speech recordings typically contain responses to elicited speech tasks rather than the unstructured connected speech typically used in identification experiments. Some speech task responses do contain connected speech (e.g., paragraph reading), but others are very dissimilar (e.g., vowel prolongation). The impact of speech task on identifiability remains unknown.

In this study we address the risk of re-identification in a series of experiments exploring the re-identifiability of medical speech recordings without using any metadata. We accomplish this goal by modeling an adversarial attack using a state-of-the-art speaker identification architecture, wherein an adversary trains the speaker identification model on publicly available, identified recordings and applies it to a set of unidentified clinical recordings. We first demonstrate that the risk is inversely related to the number of comparisons an adversary must consider, with this search space being a function of the size of the shared dataset and the known-speaker dataset used by the adversary. The risk is high for a small search space but drops sharply as the search space grows. Next, we show that the nature of a speech recording substantially influences re-identification risk, with recordings containing non-connected speech (e.g., vowel prolongation) being harder to identify. Our findings suggest that speaker identification models can be used to re-identify participants in principle, but in practice, the risk of re-identification appears small when the search space is large.

## Methods

Our experimental design was premised on the following assumptions: (1) a data recipient has decided to attempt re-identification of study participant data, therefore becoming an adversary, and (2) this adversary relies on an adversarial attack strategy known as a marketer attack, wherein they use a large dataset of identified speech (hereafter the 'known' set), perhaps obtained from an online source such as YouTube, to train a speaker identification model that is then used to re-identify as many unknown speakers in the shared clinical dataset (hereafter the 'unknown' set) as possible [19, 31]. Other attack scenarios are possible, but a marketer attack establishes an accepted baseline for risk. To simulate this attack scenario, we built a text-independent speaker identification model with a combination of x-vector extraction using ECAPA-TDNN [32] and a downstream PLDA-based classifier [33, 34]. Figure 1 depicts the architecture of our model.

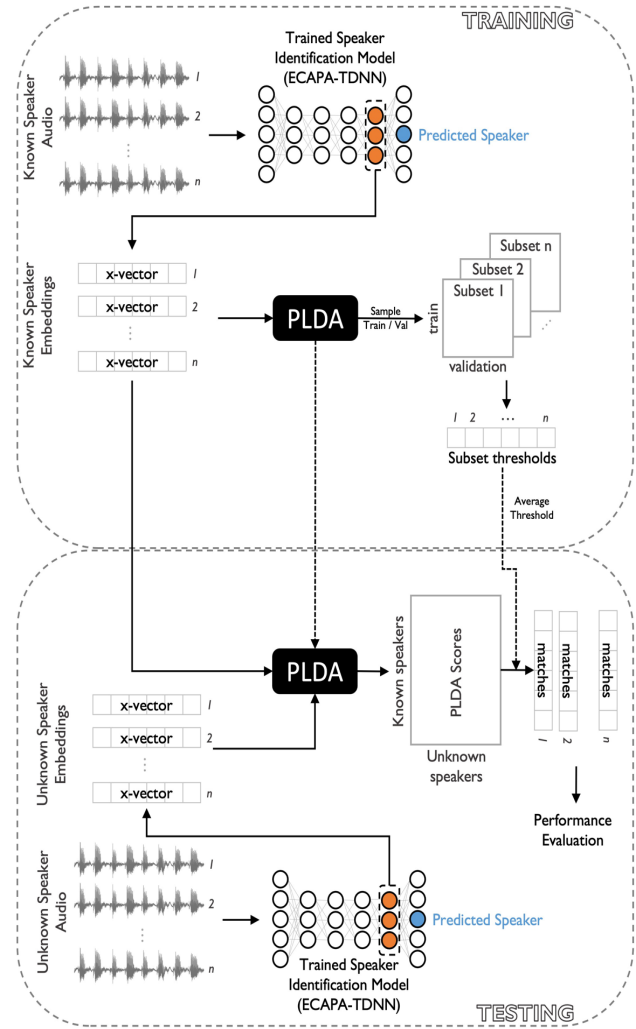

*Figure 1. Speaker identification system architecture*. During training recordings from known speakers are fed into a pretrained speaker identification model to extract embeddings. These represent a low dimensional, latent representation for each recording that is enriched for speaker-identifying features (x-vectors). We used these x-vectors for known speakers to train a PLDA classifier and generate an average threshold for acceptance/rejection of a speaker match over several subsets. During testing, the extracted x-vectors are fed into the trained PLDA and the training threshold is applied, resulting in a set of matches (or no matches) for each recording.

Data

An ideal dataset for our attack scenario would consist of (1) a set of elicited speech recordings from tasks typically used in clinical or research speech evaluations, and (2) a set of unstructured speech recordings including the same speakers as (1) but acquired at a different time and place. This would allow us to directly assess the risk of reidentification of medical recordings by training a model on unstructured connected speech, like what an adversary may find online. Such a dataset does not exist. As such, we made use of two separate datasets. The first was a combination of the well-known VoxCeleb 1 and 2 datasets, which contain recordings from an online source for over 7000 speakers [35, 36, 23]. The second was a medical speech dataset from Mayo Clinic, which contains recordings of commonly used elicited speech tasks but with fewer speakers.

*VoxCeleb*

The VoxCeleb 1 and 2 datasets are recent large-scale speaker identification datasets containing speech clips extracted from celebrity interviews on YouTube [35, 36, 23]. The utterances are examples of natural, real-world speech recorded under variable conditions for speakers with different ages, accents, and ethnicities. VoxCeleb 1 and 2 have a combined total of 1,281,762 recordings from 7,363 speakers.

*Mayo Clinic speech recordings*

The Mayo Clinic clinical speech dataset consists of recordings from elicited speech tasks in previously recorded speech assessments. Each speaker has some combination of clips from various tasks commonly used in a clinical speech evaluation, including sentence repetition, word repetition, paragraph reading, alternating motion rates (AMRs), sequential motion rates (SMRs), and vowel prolongation [11]. Clips from speakers vary in recording medium (cassette recording vs. DVD), microphone, degree of background noise, and presence and severity of motor speech disorder(s). There are 19,196 recordings from 941 speakers (breakdown in Table 1).

*Table 1. Breakdown of number of recordings and speakers for each task in the Mayo Clinical Speech dataset.*

| Vowel Prolongation | Alternating Motion Rates (AMR) | Sequential Motion Rates (SMR) | Word Repetition | Sentence Repetition | Reading Passage |
|---|---|---|---|---|---|
| 'aaaaaah' 1734 recordings 812 speakers | 'puh', 'tuh', 'kuh' 3921 recordings 777 speakers | 'puh-tuh-kuh' 1049 recordings 564 speakers | 'catapult', 'catastrophe' 124 recordings 62 speakers[1]  *other words* 4012 samples 354 speakers[1] | 'My physician...' 238 recordings 222 speakers[2]  *other sentences* 7505 samples 551 speakers[2] | 'You wish to know...' 612 recordings 501 speakers |

[1] 354 total unique speakers  [2] 551 total unique speakers

### X-vector extraction with ECAPA-TDNN

We generated speaker embeddings using a Deep Neural Network to extract fixed-length embedding vectors (x-vectors) from speech recordings [34, 32]. This technique has been shown to outperform previous embedding techniques, such as i-vectors [37, 38], while offering competitive performance to newer end-to-end Deep Learning approaches [21, 22]. Our network of choice was the state-of-the-art ECAPA-TDNN model, pre-trained on a speaker identification task using VoxCeleb 1 and 2 [32]. This model extracts a 192-dimensional x-vector for each speech recording. The model is publicly available through SpeechBrain, an open-source AI speech toolkit [39] and is hosted on HuggingFace.

### PLDA backend classifier

Probabilistic linear discriminant analysis (PLDA) classifiers are a standard approach for speaker identification due to their ability to reliably extract speaker-specific information from an embedding space using both within- and between-speaker variance [33, 40]. PLDA is a dimensionality reduction technique that projects data to a lower dimensional space where different classes are maximally separated (i.e., maximal between-class-covariance). The advantage of PLDA over standard LDA is that it can generalize to unseen cases [41]. PLDA can then be used to decide if two datapoints belong to the same class by projecting two datapoints to the latent space and using the distance between them as a measure of similarity. This works well for speaker identification as speaker embeddings are typically fed into a classifier in pairs, where the classifier's role is to optimally reject or accept the hypothesis that the two recordings are from the same speaker. PLDA typically uses the log likelihood ratio (probability of recordings belonging to the same class vs different classes) to measure similarity, commonly referred to as PLDA scores. During training of a PLDA classifier, PLDA scores for each pairwise comparison in the training set are computed and then used to set a threshold for determining potential speaker matches [33, 40].

Our classifier was built and trained on a set of x-vectors extracted from either VoxCeleb speech recordings or Mayo Clinic speech recordings using functions from SpeechBrain [39]. We aimed to maximize performance by giving the model multiple speech embeddings per speaker during training, each extracted from recordings under different degradation conditions (varying background noise, microphone distances, etc.), that were then averaged to create a single speaker embedding [33].

### Threshold calculation for acceptance/rejection

During training, an optimal threshold needs to be determined to classify whether a given PLDA score represents a match, which can then be applied to new, unseen recordings. Generally, the Equal Error Rate (EER) is used to select the threshold [33, 21, 22, 34, 24]. The use of EER assumes that the 'cost' of a false acceptance (FA) is the same as a false rejection, such that the optimal threshold is one where the False Acceptance Rate (FAR) equals the False Rejection Rate (FRR) [22]. While this may be feasible for smaller datasets, when there are several million comparisons, the EER often generates many potential matches per speaker. As such, this can overwhelm the model early on and make it difficult for an adversary to find reliable matches. To scale to large numbers of comparisons, the adversary must make decisions on how to calibrate the threshold calculation, such as penalizing FAs more heavily even if some true acceptances (TA) are missed. From an adversary's perspective, it is less costly to miss TAs if the identified accepted cases have a high likelihood of being true. In effect, precision is more important than recall. The Detection Cost Function (DCF, Equation 1, LeeuwenBrummer07) captures this well:

$$minDCF = C_{FR} * FR * prior_{target} + C_{FA} * FA * (1 - prior_{target}) \qquad (1)$$

*Equation 1. Detection Cost Function (DCF).* We take the cost of a false rejection ($C_{FR}$) multiplied by the total number of false rejections (FR) and the prior probability of the target and add it to the cost of a false acceptance ($C_{FA}$) multiplied by the total number of false acceptances (FA) and the complement of the prior probability.

Using this function, a threshold can be found by setting optimal cost and prior terms based on the adversary's perspective (i.e., avoid FAs more aggressively) and then finding the FA and FR values that minimize the DCF (minimized decision cost function, minDCF) [42]. For example, as the prior probability of the target is lowered (i.e., if an adversary expects a small overlap), the calculation puts more emphasis on avoiding FAs (lower FAR) as compared to the EER. Increasing the cost of FAs and decreasing the cost of FRs further avoids FAs.

We used minDCF with two parameter configurations: (1) the default configuration for the SpeechBrain implementation of minDCF, where FAs and FRs are penalized equally ($C_{FA}$=1, $C_{FR}$=1, $prior$=0.01) [39] and (2) a strict configuration with a higher penalty for FAs ($C_{FA}$=10, $C_{FR}$=0.1, $prior$=0.001).

Due to the large amount of training data in VoxCeleb, it was not computationally feasible to select a threshold for the entire identified set at once. As such, we calculated minDCF multiple times on subsets of speakers and averaged across runs to generate the optimal threshold for the entire known speaker set. For each run, latent representations from two random subsets of 100 speakers were selected from the known speaker set and fed to minDCF to select a threshold. If the two subsets had no overlapping speakers, the entire run was discarded as a threshold cannot be calculated. We ran this process between 100 to 500 times depending on the overall number of speakers in the known set. Smaller known sets required fewer runs to converge on an optimal threshold.

Generating experimental speaker sets

To model the attack scenario, we randomly sampled our datasets to generate the following speaker subsets:

- *Known set:* This set represents speakers with identified audio data from an online source that the adversary has access to.

- *Unknown only set:* This set represents speakers in a shared dataset who do not have identifiable audio online. No unknown only speakers are present in the known set.

- *Overlap set:* This set is a proxy for speakers in a shared dataset who do have identifiable audio somewhere online. Some speakers from the known set are randomly selected to create this set.

- *Unknown set:* This represents the full shared dataset, consisting of both the unknown only set and overlap set.

The number of speakers per set varied based on the experiment. Furthermore, the number of speech recordings per speaker varied in the known and unknown set. We used all available speech recordings per speaker in the known set but randomly selected only one recording per speaker in the unknown set. For overlapping speakers, the selected recording for the unknown set was withheld from the known set. The limit of one sample per speaker in the unknown set is based on the nature of a supposed real-world dataset where all speech is unlinked and partially de-identified, meaning the adversary needs to separately find potential matches for each recording even if they come from the same speaker.

Because we randomly subsampled speakers to generate these sets, there is variation in the selected speakers selected for each experiment, which will result in variability in model performance that is dependent only on the dataset. To account for this, we generated multiple speaker splits per experiment. The exact number of splits was dependent on the experiment.

Experiments

*VoxCeleb realistic experiments: Effect of search space size*

We relied upon VoxCeleb 1 & 2 to investigate the capability of an attack as a function of the size of the search space (i.e., the number of comparisons made to find matching speakers). We re-identified speakers by comparing each speaker in the known set to each speaker in the unknown set. The search space is thus the product of the size of the known and unknown set. As such, an increase in either set will increase the number of comparisons. We consider both cases separately, which allows us to consider one scenario that is dependent on the resources of the adversary (known set size) and another that is under the control of the sharing organization (unknown set size).

To construct a realistic scenario, we assumed that the known and unknown set would have a low amount of speaker overlap. To justify this assumption, one can consider what would be involved in constructing a set of known speakers. In the absence of metadata about the unknown speakers (e.g., the ages, location, etc.), there would be no way for an adversary to target a specific population to build out their known set. It is unlikely to be feasible for an adversary to manually collect and label speech recordings for a large proportion of the population. Instead, an adversary would likely need to rely on a programmatic approach using easily accessible identifiable audio, such as scraping audio from social media and video or audio sharing websites [43]. It is worth noting that this would still be difficult because of several confounding factors: (1) not all members of the population use these websites; (2) not all users have publicly accessible accounts; (3) users with publicly accessible accounts may not have identifiable information linked to it; (4) some accounts post audio/video from multiple speakers, including speakers who also have their own accounts; (5) many users do not post at all; and (6) the population of users is not representative of the general US population, let alone the subset with speech disorders – in terms of the distributions of both age and geographic area [44]. As such, there is no reason to suspect that a patient in a shared medical speech dataset would have a high likelihood of existing in an adversary's set of identified audio recordings.

We also assumed that the adversary would not know which unknown speakers, if any, exist in the known speaker set. The adversary must therefore consider all potential matches rather than only focusing on the $N$ overall best matches, where $N$ is the known overlap. This would reduce the reliability of any match since the likelihood of all potential matches being true is lower than the likelihood of the best $N$ matches being true.

We first trained the speaker identification model with the numbers of speakers in the known set increasing from 1000 to 7205 while maintaining a static unknown set size of 163 speakers with low speaker overlap between sets (5 speakers in the overlap set, 158 in unknown only set).

We then trained the model with a fixed known set size of 6000 speakers while increasing the number of speakers in the unknown set from 150 to 1000 speakers and maintaining a low overlap of 5 speakers.

Given the low number of overlapping speakers and overall large set sizes, we generated 50 speaker splits for each set size of interest (known set: 1000, 4000, 7205; unknown set: 150, 500, 1000).

The accept/reject threshold for these experiments was set using the strict minimized decision cost function (minDCF) configuration.

*VoxCeleb known overlap and full overlap experiments: Worst-case scenarios*

There are two important initial assumptions in our construction of realistic experiments: (1) the adversary was unaware of the amount of overlap between known and unknown sets and (2) the amount of overlap was low. Thus, we considered how re-identification risk would be affected if either assumption was incorrect.

First, we considered a potential worst-case scenario in which the adversary did know the amount of overlap speakers $N$ and was therefore able to limit potential matches to the top $N$ best matches. As previously mentioned, limiting the number of matches could theoretically improve model reliability, and further reducing the number of matches could produce more noticeable effects. We leveraged our base results from the realistic experiments and only considered the top $N$ best matches.

Next, we considered a less realistic worst-case scenario wherein all unknown speakers exist in the known speaker set. From an adversary's perspective, a full overlap scenario would provide the best chance for the adversary to successfully re-identify speakers since most FAs occur when the model finds a match for unknown speakers who are not in the known speaker set.

We assessed this scenario by replicating the realistic experiments with full overlap between known and unknown sets. That is, regardless of the unknown set size, all speakers also exist in the known set (no unknown only set). When increasing the known set size with a fixed unknown set of 163 speakers, the overlap set consists of all 163 speakers, and when increasing the unknown set size with fixed unknown set, the overlap set is the same as the unknown set size of interest (150, 500, 1000). In this scenario, we generated only 20 speaker splits for each set size of interest as the larger overlap set led to less variance across runs.

As in the realistic experiments, the accept/reject threshold was set using the strict minDCF configuration.

*Mayo Clinic speech recordings experiments: Effect of speech task*

Next, we shifted our focus from the public VoxCeleb dataset to a private dataset of Mayo Clinic medical speech recordings to look at factors specific to a clinical speech dataset, such as whether certain elicited tasks are easier for re-identification and whether being able to link recordings to the same speaker across tasks (pooling) increases risk.

We first compared the performance of the speaker identification model across the various elicited speech tasks in the Mayo dataset based on the same adversarial attack scenario used with the VoxCeleb experiments. In this scenario, the cross-task performance aligns with a real-world case where the training data contains connected speech recordings (i.e., recordings of continuous sequences of sounds like that of spoken language), but speakers are reidentified using a variety of elicited speech tasks (Table 1). Each task has a different degree of similarity to connected speech (left: most, right: least):

*Reading Passage > Sentence Repetition > Word Repetition > SMR > AMR > Vowel Prolongation*

The reading passage is essentially real-world connected speech in terms of content and duration, but sentence repetition is closer to the connected speech seen in most speech datasets [23]. As such, we selected sentence repetition recordings for speakers in the known set.

The resulting known set consisted of 500 speakers and included all sentence repetition recordings excluding any repetitions of the physician sentence ('My physician wrote out a prescription') which was saved for the unknown set. We then generated separate unknown sets for each elicited task with 55 speakers (5 overlap, 50 unknown only) who had both sentence repetition recordings and a recording for the given re-identification task (e.g., 'My physician...' sentence, alternating motion rates, etc.).

The known and unknown set sizes were bounded by the number of speakers with sentence repetition recordings (587 speakers), as the sentence-sentence configuration needed enough speakers to create a separate known and unknown only set. We also considered the sentence-sentence configuration (i.e., sentence repetitions in both the known and unknown set) as the realistic baseline.

As a secondary part of this experiment, we pooled all available recordings from all elicited speech tasks (by averaging their embeddings) to generate an unknown set where the adversary can link recordings from a given speaker, i.e., there would be more speech for each unknown speaker.

In addition to the cross-task performance, we also compared the within-task performance — where the same elicited speech task is used for both known and unknown speakers — to determine if anything about the nature of a given speech task affects re-identification. For example, the variance across recordings for the sentence repetition task reflects a combination of static speaker factors (e.g., identity, and age), dynamic speaker factors (prosody, e.g., the same speaker may emphasize different words in a sentence on repeated trials), and content factors (i.e., different words in different sentences). By contrast, a task like alternating motion rates involves repeating the same syllable as regularly and rapidly as possible, with most of the variance across speakers likely resulting from static speaker factors. A priori, considering all the elicited tasks, one would expect the proportion of variance across speakers due to dynamic speaker factors to decrease following the same scale as similarity to natural speech. The reading passage would have the most variance due to dynamic speaker factors alone while vowel prolongation would have the least. By removing the confounding variable of different elicited tasks for known and unknown speakers (i.e., the model is both trained and tested on the same task), we can ascertain whether qualities of the speech task itself influence re-identification.

We used the same set sizes as the cross-task experiments (500 known, 55 unknown, 5 overlap) but used recordings from the same elicited speech task in both the known and unknown set. This setup required at least two recordings per speaker for each task. Some tasks had fewer than 500 unique speakers or not enough recordings (word repetition, reading passage), so not every known set had exactly 500 speakers. The word repetition task had 299 speakers and the reading passage task had 466 speakers.

To account for the decrease in amount of data as compared to VoxCeleb experiments, we generated only 20 speaker splits per task with default minDCF parameters.

### Statistical Analyses

Our primary outcome of interest was the average number of false acceptances (FAs), where the model accepts a match for an unknown speaker without a true match, as compared to true acceptances (TAs) over several subsampled datasets. This informed the reliability of re-identification. Using the counts, we also calculated the Pearson's correlation coefficient between FAs and set size along with the false acceptance rate (FAR) to determine if a linear correlation existed between the number of FAs and number of speakers/comparisons. A t-test was performed to find the significance of each correlation.

## Results

### VoxCeleb realistic experiments: Effect of search space size

When training the speaker identification model with increasing numbers of speakers in the known set while maintaining a static unknown set size with low speaker overlap between sets, we found that increasing the number of speakers in the known set resulted in an increase in the mean number of FAs while TAs remained stable, with a linear correlation between FAs and number of known speakers ($r = 0.30$, $P < .001$, $t_{148} = 3.89$, Figure 2a). Increasing the size of the unknown set had a similar yet more pronounced effect than increasing the known set size, with a higher linear correlation between FAs and number of unknown speakers ($r = 0.60$, $P < .001$, $t_{148} = 9.21$, Figure 2b).

The difference in effect can be understood based on the geometry of the search space. While the unknown set remains significantly smaller than the known set, adding any additional speaker to the unknown set will result in a larger increase in the search space than adding a speaker to the known set. As such, we can better demonstrate the overall trend in FAs by considering the results in terms of total comparisons (i.e., search space size) rather than individual set size.

We observed that there was a high positive linear correlation between FAs and the number of comparisons ($r = 0.69$, $P < .001$, $t_{198} = 13.54$, Figure 2c), with the mean FAs increasing from 0.04 to 2.84 while TAs remained stable. There was a corresponding drop in precision (Figure 3a). It was notable that the FAR remained low and relatively stable, averaging at $4.152 * 10^{-07}$ (Figure 3b), indicating that the demonstrated trend should hold for the larger numbers of comparisons we would expect to see in a real attack.

We further observed that using a stricter threshold for matches resulted in our model selecting only one match per speaker. This is functionally the same as limiting matches to only the best potential match for each speaker (rank 1 matches), which is an option for an adversary to increase reliability without knowledge of the amount of overlap.

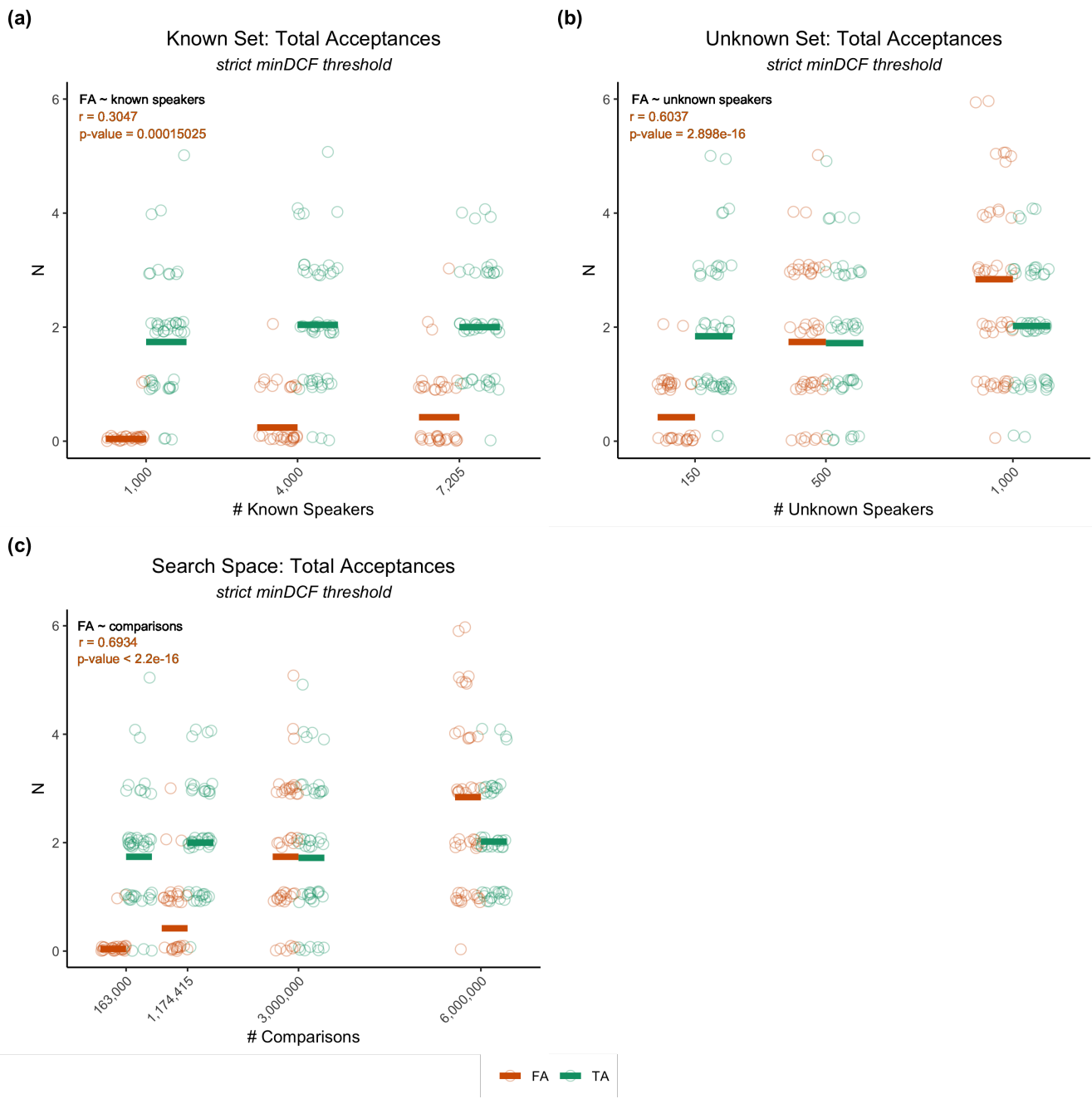

Figure 2. Number of true and false acceptances for the speaker recognition model in a realistic scenario with VoxCeleb. *(a)* shows the counts when varying the number of known speakers while keeping the number of unknown speakers static, *(b)* shows the counts when varying the number of unknown speakers while keeping the number of known speakers static, and *(c)* shows the overall trend in terms of number of comparisons made (i.e., the search space size = known ∗ unknown speakers). All plots *(a-c)* include the Pearson's correlation coefficient and corresponding significance for false acceptances and number of speakers/comparisons. Each run is plotted as a single circle, with red horizontal lines indicating the mean number of false acceptances and green horizontal lines indicating the mean number of true acceptances.

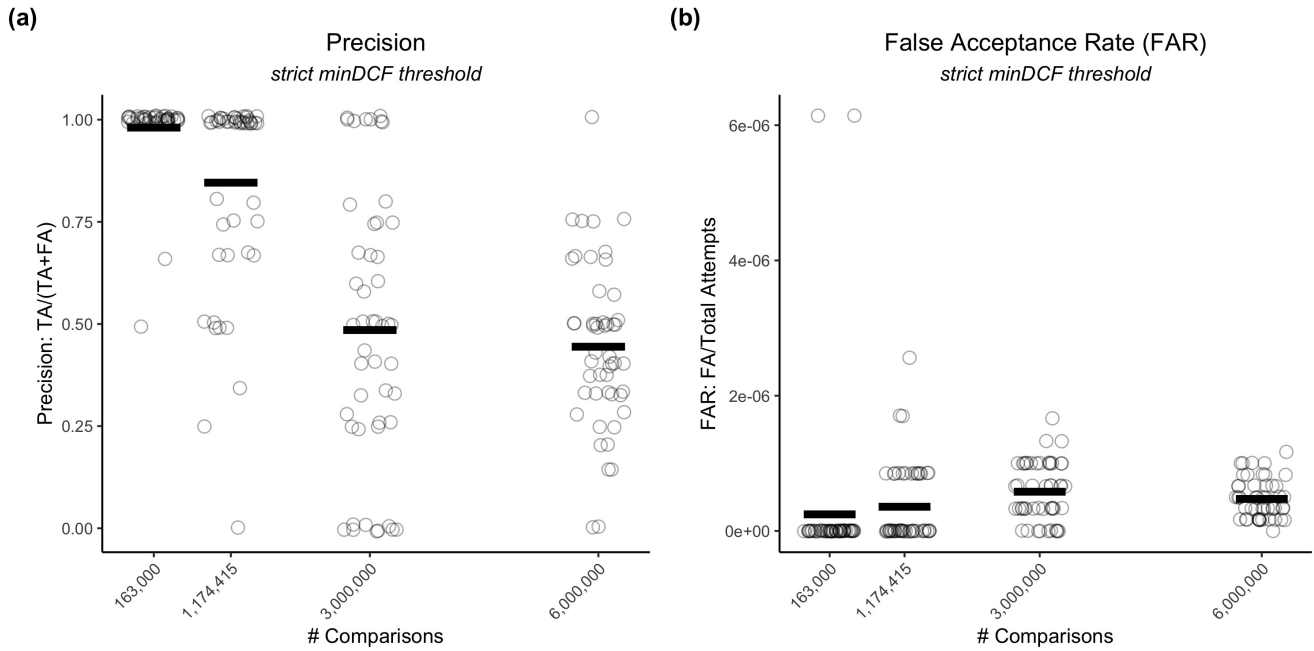

*Figure 3. Precision and False Acceptance Rates for the speaker recognition model in a realistic scenario with VoxCeleb. Precision (a) and false acceptance rates (b) are shown as a function of the number of comparisons. For both plots, each run is represented by a circle, and the mean is represented by a horizontal black line.*

### VoxCeleb known overlap and full overlap experiments: Worst-case scenarios

When only considered the top *N* best matches, we found that there was still a trend of increasing FAs, with a high linear correlation to number of comparisons ($r$ =0.70, $P$ < .001, $t_{198}$ =13.72, Figure 4a). This indicates that some FAs were seen as better matches than some TAs, as further supported by the associated drop in precision (Figure 4b).

When all unknown speakers exist in the known speaker set, the performance improved significantly, with most matches being correct (Figure 4c). Even so, there was still a high positive linear trend for FAs, indicating that at high overlap, some FAs were ranked higher than true acceptances ($r$ =0.67, $P$ < .001, $t_{78}$ =7.98, Figure 4d). This is surprising given that for the realistic experiments, all FAs were associated with matches for non-overlapping speakers.

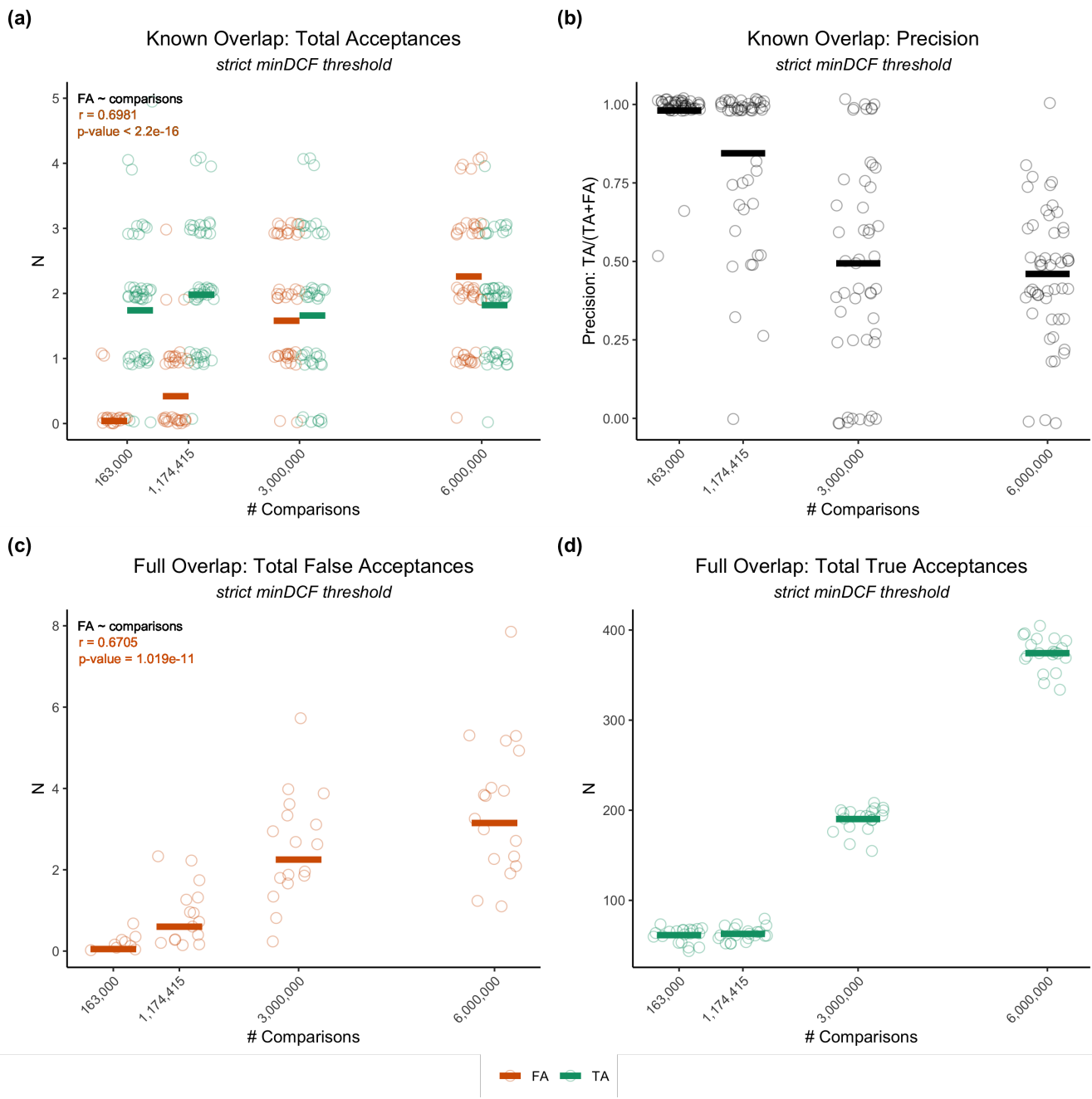

*Figure 4. Results for our speaker recognition model in worst-case scenarios with VoxCeleb. (a)* shows the true and false acceptance counts for a known overlap scenario (limited to *N* = 5 best matches), while *(b)* shows the corresponding precision as a function of the number of comparisons (search space size). *(c)* and *(d)* show the false and true acceptance counts for a full overlap scenario, where all unknown speakers are present in the known speaker set as a function of the number of comparisons (search space size). *(a)* and *(c)* also shows the Pearson's correlation coefficient and corresponding significance between false acceptances and number of comparisons. Each run is plotted as a single circle, with red horizontal lines indicating the mean number of false acceptances, green horizontal lines indicating the mean number of true acceptances, and black horizontal lines indicating the mean precision.

## Mayo Clinic speech recordings experiments: Effect of speech task

We first compared the performance of the speaker identification model across the various elicited speech tasks in the Mayo dataset based on the same adversarial attack scenario used with the VoxCeleb experiments. We observed that the total number of acceptances decreased as the unknown speaker tasks became less like the known speaker task, but the proportion of TAs and FAs also varied. This made it more difficult to determine the performance through counts alone (Figure 5a). When considering precision instead, we found that the baseline (sentence-sentence) had the best performance though average precision was not high (66.5%, Figure 5b). The reading task, the word repetition and sequential motion rate (SMR) tasks had comparable performance. However, alternating motion rate (AMR) and vowel prolongation tasks had extremely low precision. Vowel prolongation had a precision of zero (no TAs) but a high number of FAs. Pooling resulted in decreased performance as compared to the baseline and the top performing tasks, likely due to the influence of AMR and vowel prolongation recordings.

The within-task results did not exhibit the same effect as the cross-task. We found that all tasks re-identified the overlapping speakers (TA=10), but the number of FAs varies drastically across task (Figure 5c). Whereas before the baseline had the best performance, we instead see that SMR and vowel prolongation tasks have the highest precision (Figure 5d). In fact, as tasks became more dissimilar from connected speech and had less variance due to dynamic speaker factors, they saw a relative increase in performance as compared to the cross-task scenario.

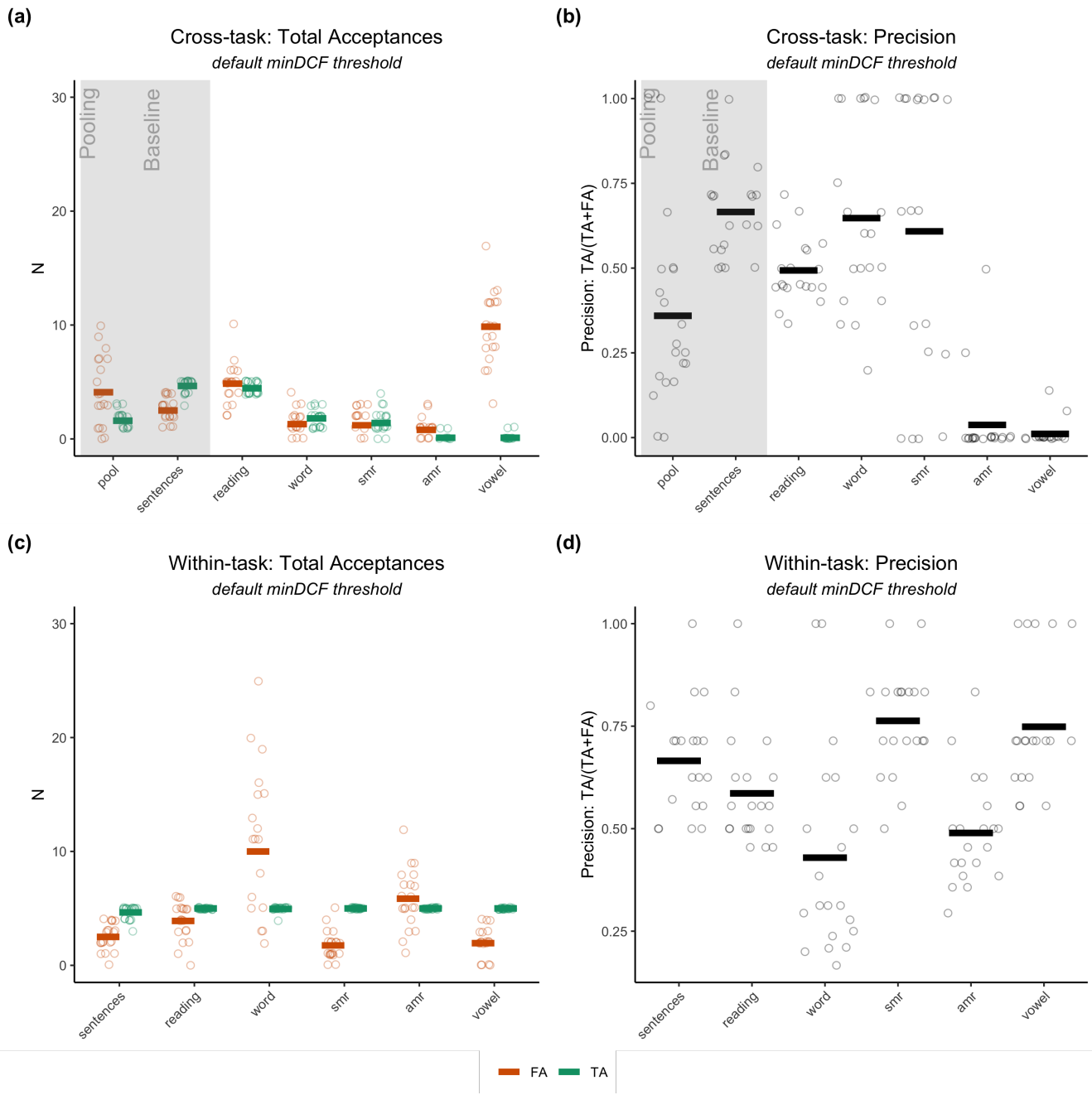

*Figure 5. Results for our speaker recognition model with the Mayo clinical speech dataset. (a)* and *(b)* show cross-task results, where recordings for known speakers are always sentence repetition but the task for unknown speaker recordings varies. The baseline is when sentence repetitions are in both the known and unknown set. Pooling is when all recordings for an unknown speaker are linked together across all tasks. *(a)* show the breakdown of counts for this case while *(b)* is the corresponding precision. *(c)* and *(d)* show within-task results, where tasks for known and unknown speakers are always the same. *(c)* is the breakdown of counts for this case while *(d)* is the corresponding precision. Each run is plotted as a single circle, with red horizontal lines indicating the mean number of false acceptances, green horizontal lines indicating the mean number of true acceptances, and black horizontal lines indicating mean precision.

# Discussion

## Principal Results

In this study, we investigated the risk of re-identification of unidentified speech recordings without any other speaker or recording related metadata. To do so, we performed a series of experiments reflecting a marketer attack by an adversary with access to identified recordings from a large set of speakers and the capability to train a speaker identification model, which would then be used to re-identify unknown speakers in a shared dataset. We systematically considered how changes in the size of the datasets and the nature of the speech recordings affect this risk of re-identification. We found that it is feasible to use a speaker identification design – a deep learning speaker embedding extractor (x-vectors) coupled with a PLDA backend – to re-identify speakers in an unknown set of recordings by matching them to recordings from a set from known speakers. Given the performance of current state of the art speaker identification models, this is not surprising. However, these models have only rarely been applied in an adversarial attack scenario [24, 25] – i.e., their potential as an attack tool for an adversary who aims to re-identify speakers in a shared or publicly available dataset was largely unknown. Furthermore, the feasibility of such an attack has not been considered and may have been assumed to be low for speech recordings stripped of all metadata (sometimes referred to as de-identified or anonymous in the literature) without considering the identifiability of the acoustic signal itself [45, 46, 47, 48].

Our findings suggest this is not true. Consistent with a previous study that found high re-identification risk for one unknown speaker with known sets of up to 250 speakers (search space <= 250 comparisons) [25], we observed that risk was indeed high for small search spaces. For example, when attempting to re-identify 5 overlapping speakers between a small set of unknown speakers (n=163) and a moderate set of known speakers (n=1000) our model had nearly perfect precision (Figure 3a) and identified 2 speakers on average (Figure 2a). However, our experiments allowed us to extend this to more realistic search spaces, such as scenarios where an adversary uses a known speaker set of up to 7205 speakers and an unknown speaker set of up to 1000 speakers (search space <= 6,000,000 comparisons). We observed that the risk dropped sharply as the search space grew. The FAR was relatively stable at $4.152 \times 10^{-7}$ (Figure 3b), which translates to an average of increase of one FA for every 2.5 million comparisons. This is a key take home message from these experiments – increasing the size of the search space, whether through growing the size of the adversary's set of identified recordings or the shared dataset, resulted in a corresponding increase in the number of FAs. Given that the number of overlapping speakers remained constant, this suggests that the primary driver of FAs is the size of the non-overlapping known to unknown comparison space, i.e., most FAs arise from non-overlapping unknown speakers being falsely matched to known speakers. In fact, all FAs in the realistic experiments correspond to non-overlapping unknown speakers. Here, it is worth noting that in the experiments where we only considered the top N matches (where N = number of overlapping speakers), this trend remained true since some of the FAs were scored higher than TAs (Figure 4). This suggests that for a sufficiently large search space, even considering only the best N matches will result in many FAs. We pushed this line of reasoning to its limit by considering a worst-case scenario of full overlap, where all unknown speakers had a true match. Even in this scenario there were still many FAs, and the proportion of FAs grew with increasing search space size. Importantly, this scenario showed that overlapping speakers can still be falsely matched when the overlap is high.

Our experiments with the Mayo clinical speech recordings allowed us to assess the influence of speech task based on both cross-task and within-task performance. When the model was trained on sentence repetition (i.e., the known dataset consists of sentence recordings) and then applied to other tasks (i.e., the unknown set consists of elicited, non-sentence speech), the precision slightly improved for the reading passage, word repetition, and SMR tasks but deteriorated for the less connected speech-like tasks such as AMR and vowel. These results can be understood with reference to the default minDCF settings, which would penalize FAs and false rejections (FRs) equally. The threshold was chosen using sentence repetition task recordings, such that in most instances, all overlapping speakers were re-identified for unknown sets with connected speech tasks (sentence repetition, reading, word repetition, SMRs). The minDCF threshold for these similar tasks resulted in fewer overall acceptances (higher false rejection rate (FRR)), but as the tasks diverged from sentence repetition with respect to the degree of connectedness, they were also less likely to be FAs. This suggests that identifiable characteristics learned from training on the sentence repetition task translate well to other connected speech tasks. It also demonstrates the difficulty of choosing a threshold when the tasks in the known set are different from the unknown set – because of difference within a speaker across tasks, it becomes hard to balance the goal of some TAs with a flood of FAs as search space increases. In this instance, a slightly stricter threshold may have been better for the adversary. By contrast, the non-connected speech tasks (AMRs, vowel prolongation) had almost no TAs and a high number of FAs, suggesting that identifiable characteristics from connected speech tasks do not translate to non-connected speech tasks. This is not unexpected given that models perform worse when tested on data that is dissimilar from the training data [49, 50]. Following this, we also found that pooling across tasks decreased performance from the baseline. Generally, having more data for a speaker is expected to improve performance, but it is possible that adding recordings of non-sentence tasks to the unknown set hurt performance because the identifiable characteristics are different across tasks and the system is unable to accommodate. In other words, any helpful characteristics from the connected speech tasks were cancelled out by competing characteristics from the non-connected speech tasks.

In the within-task scenarios, where the known and unknown set were made up of the same task, the reidentification power for overlapping speakers was better than in the cross-task scenario, but tasks exhibited vastly different FA rates. In fact, many tasks that were different from connected speech saw improved performance. For example, vowel prolongation, which is non-connected and the most perceptually different from sentence repetition, exhibited the worst cross-task performance but the second best within-task performance. This may be that due to fewer interfering dynamic speaker factors, this specific task isolated acoustic features that are tied to identity well, which allows for good within-task performance.

Another important finding is that performance for sentence repetition was much weaker than expected based on the VoxCeleb experiments with a larger number of comparisons. We suspect this may be due to a combination of factors. First, it may be more difficult to differentiate speakers in an unknown set of elicited recordings where every speaker is uttering the same sentence. Second, the clinical recordings were all made by patients referred for a speech examination. Consequently, the resulting cohort contains mostly abnormal speech, which may impact the PLDA performance. Third, the Mayo clinical speech dataset is smaller than the VoxCeleb dataset both in terms of number of speakers and number of recordings per speaker, and the recordings are also shorter in duration. This likely had a negative impact on the training of the PLDA classification backend. It remains unknown if larger clinical datasets or datasets with more recordings per speaker may yield findings more like the VoxCeleb results.

Taken together, our findings suggest that the risk of re-identification for a set of clinical speech recordings devoid of any metadata in an attack scenario like the one we considered here is influenced by (1) the number of comparisons an adversary must consider, which is a function of both the size of the unknown dataset and the known datasets; (2) the similarity between the tasks and/or recordings in the unknown and the known datasets; and (3) the characteristics of the recordings in the unknown dataset, such as degree of speaker variance and presence and type of speech disorders. These findings translate to actionable goals for both an adversary and the sharing organization.

### Mitigating privacy risk

While we assumed that the sharing organization already reduced risk by stripping recordings of demographic (e.g., age or gender) or recording metadata (e.g., date or location), we additionally suggest that re-identification risks could be further reduced by increasing the search space (i.e., larger shared dataset size) or decreasing the similarity between shared recordings and publicly available recordings (e.g., sharing vowel prolongation as long as a publicly available vowel prolongation dataset does not exist, or sharing a larger variety of speech disorders instead of a single disorder). Even if the number of overlapping speakers increased with the size of the shared dataset, the results from the full-overlap scenario indicate that a model could still have reduced reliability due to an increasing FAR.

By contrast, the adversary can also use this knowledge to enhance their attacks. From their perspective, any additional information that can reduce the search space or increase similarity between recordings will increase the reliability of speaker matches. This could involve using demographics such as gender, be it shared or predicted by a separate model, to rapidly reduce the number of comparisons. For instance, when gender balance is 50:50, comparing unknown males to known males would reduce the number of comparisons by 75%. The adversary may also seek out publicly available recordings of abnormal speech to refine their model(s) or reduce the search space. If social media groups exist where identified users with certain medical or speech disorders post video or audio, an adversary could restrict their known set to these users. Similarly, research participants and support staff may also influence risk through disclosure of participation. By disclosing participation in a study known to share speech recordings, a participant would effectively reduce the size of the known set to one, increasing their individual risk of re-identification. In addition, having a confirmed match can increase risk overall as the adversary would have a baseline to determine the reliability of matches [51]. Although the focus of this investigation has been on the change in relative risk with changes in dataset size and speech task, it is worth considering our findings in the context of other factors that impact risk in practice. The most obvious is the availability of additional data on the speakers or recording metadata. In this respect, it is worth noting that sufficient demographic data even in the absence of speech is well known to carry a significant risk of re-identification [19, 52]. If any aspect of the metadata makes a patient population unique (i.e., there is only one person in a given age range), the risk of re-identification increases [12, 14]. Furthermore, if any data about the speaker (e.g., sex) or recording (e.g., date) reduces the search space, re-identification risk would increase. There may also be identifiable content in the recording. During less structured speech tasks such as recordings of open-ended conversations, participants may disclose identifiable information about themselves (e.g., participant saying where they live). Removing these spoken identifiers is an active area of research [25].

Still, it is important to acknowledge that simply because records are vulnerable to re-identification does not mean that they would be re-identified. Notably, when assessing privacy concerns, the probability of re-identification during an attack is conditional on the probability of an attack occurring in the first place [52]. In most instances where data is shared, the receiving organization or individual will not have any incentive to attempt re-identification. The sharing organization, and in some cases a receiving organization, may also take steps to discourage the risk of an attack. These may take the form of legal (e.g., data sharing agreements) or technical (e.g., limited, monitored access) deterrents to a re-identification attack [53]. By contrast, the risk of an attack may be higher for publicly available datasets [54], but there may also be a greater risk for re-identification without a targeted attack. For example, in the field of facial recognition some companies have scraped billions of photos from publicly available websites to create massive databases with tens of millions of unique faces. These are then used to train a matching algorithm [43] which an end user could query using a photo of an unknown face and obtain a ranked list of matching faces and the source (e.g., Facebook). The end user can visit the source website and instantly gain access to other data that may increase or decrease their confidence in a match as well as provide feedback on matches thereby gradually increasing the performance of the tool as well as the number of known faces. If similar databases are built for speech recordings, it will certainly include publicly available medical speech recordings. Every query to the model would then represent a threat to such a public sample being matched to a queried recording regardless of the intent of the user that queried the model. Such a scenario is difficult to simulate because of the continuously improving nature of the algorithm and the fact that users would incorporate various degrees of non-speech data.

Refraining from publicly releasing datasets is an obvious mitigation strategy to some of these threats. However, the risk of re-identification must always be balanced with the benefit of data sharing, since larger, more representative datasets for development and testing of digital tools may benefit patients. It is critical that policymakers consider this balance in the context of the rapidly evolving field of artificial intelligence. Naïve approaches such as the 'deidentification release-and-forget model' are unlikely to provide sufficient protection [55]. Similarly, informed consent for public release is problematic because the risk of re-identification will neither be static nor easily quantifiable over time. This has led to the development of potential alternative approaches such as data trusts, synthetic data, federated learning, and secure multiparty computation [56, 57, 58, 59].

### Limitations

It should be recognized that there are several notable limitations to our investigation. First, while we relied upon state-of-the-art learning architectures, the risk may differ if other computational approaches are considered [21, 22]. Second, we did not consider multi-stage adversarial attacks where one model is used to predict a demographic, such as sex or age, that is then used to limit the search space, or a scenario where an adversary manually goes through all potential matches to attempt manual identity verification. However, such approaches would introduce additional uncertainty for the adversary and may still require many comparisons depending on the sizes of the datasets, in which case our results would apply [60, 61]. Third, we did not directly consider the risk for normal versus abnormal speech. Nearly all recordings in the Mayo Clinic speech dataset contain abnormal speech, whereas all VoxCeleb recordings are all from normal speakers. Ideally, there would be a single dataset that contains both. Fourth, it should be noted that beyond methodological limitations, our results may not generalize well outside of the US since the VoxCeleb data has a strong US bias, and all the Mayo recordings were captured in the US. As such, it will be important to perform future experiments that leverage alternative computational architectures, more complex adversarial attacks, conversational speech, and data from other geographic regions to assess the re-identification risk for medical speech data more comprehensively.

## Conclusions

In summary, our findings suggest that while the acoustic signal alone can be used for re-identification, the practical risk of re-identification for speech recordings, including elicited recordings typically captured as part of a medical speech examination, is low with sufficiently large search spaces. This risk does vary based on the exact size of the search space — which is dependent on the number of speakers in the known and unknown sets — as well as the similarity of the speech tasks in each set This provides actionable recommendations to further increase participant privacy and considerations for policy around public release of speech recordings. Finally, we also provide ideas for future studies to extend this work, most notably the need to assess other model architectures and datasets as improvements in speaker identification could substantially increase re-identification risk.

## Data Availability

VoxCeleb 1 and 2 are publicly available at https://www.robots.ox.ac.uk/~vgg/data/voxceleb/

Our Mayo clinical speech recordings are not publicly available because of the risk related to the release of the data. Interested researchers can contact the corresponding author for case-by-case data sharing.

## Code Availability

We used python to implement our code for pre-processing, extracting speaker embeddings, generating subsampled datasets, and running the PLDA. The source code will be available at https://github.com/Neurology-AI-Program/Speech_risk. The repository will also contains detailed documentation for using the scripts.

## Author Contributions

D.A.W., B.A.M., and H.B. conceived the ideas presented in this work and validated the results. J.R.D., R.L.U., and D.T.J. provided the resources necessary for this work. D.A.W., J.L.S., and H.B. developed the methodology for the experiments. D.A.W. curated the data for the Mayo Clinic speech recordings dataset. D.A.W. and H.B. developed the code for running experiments and visualizing the results. D.A.W. ran formal statistical analysis on the data. D.A.W. and H.B. wrote the original draft of the manuscript. All authors reviewed and edited the manuscript. D.T.J. and H.B. supervised.

## Competing Interests/disclosures

D.A.W.: No disclosures

B.M.: Dr. Malin receives funding from the NIH

J.R.D.: Dr. Duffy receives funding from the NIH

R.L.U.: Dr. Utianski receives funding from the NIH

J.L.S.: Dr. Stricker receives funding from the NIH

D.T.J.: Dr. Jones receives funding from the NIH

H.B.: Dr. Botha receives funding from the NIH

# Abbreviations

AMR: Alternating Motion Rate

DCF: Decision Cost Function

ECAPA-TDNN: Emphasized Channel Attention, Propagation and Aggregation in Time Delay Neural Network

EER: Equal Error Rate

FA: False Acceptance

FAR: False Acceptance Rate

FR: False Rejection

FRR: False Rejection Rate

HIPAA: Health Insurance Portability and Accountability Act

minDCF: Minimized Decision Cost Function

NIH: National Institute of Health

PLDA: Probabilistic Linear Discriminant Analysis

SMR: Sequential Motion Rate

TA: True Acceptance

TR: True Rejection

## Supplementary information

The VoxCeleb datasets are originally split into a development set of 7205 speakers and a test set of 163 speakers. In the realistic experiments, we selected speakers randomly from the combined set of VoxCeleb 1 and 2 speakers (7,368 total speakers) (see "Methods"). We also ran a secondary set of experiments where we used only the development set when selecting speakers for the known speaker set and used all the test set as the unknown speaker set. For overlap, we selected 5 random speakers from the known set and added them to the unknown set. As a result of this design, most speakers in the unknown set were fixed across runs.

We found that there were overall more false acceptances (FAs) as compared to the corresponding realistic experiment (Figure 2a), with a higher positive correlation between FAs and number of known speakers ($r = 0.78$, $P < .001$, $t_{68} = 10.43$, Figure 6a vs. $r = 0.30$, $P < .001$, $t_{148} = 3.89$, Figure 2a). As expected, the corresponding precision also dropped much lower ($< 0.25$, Figure 6b vs. $> 0.75$, Figure 3a). In addition, the average false acceptance rate (FAR) was much higher at $\sim 5*10^{-6}$ (Figure 6c). As mentioned in the methods, the FAR may have been underestimated in our experimental setup because the ECAPA-TDNN embedding model was exposed to the 'unknown' set during its training [32]. This suggests the true FAR for the Voxceleb 1 and 2 datasets likely lie somewhere in between $\sim 5*10^{-6}$ (Figure 6c) and $\sim 4.152*10^{-7}$ (Figure 3b). However, all our conclusions are based on the lower of the FAR bounds, which is the most conservative and privacy promoting option. We present the high end of the FAR bound here to provide further context.

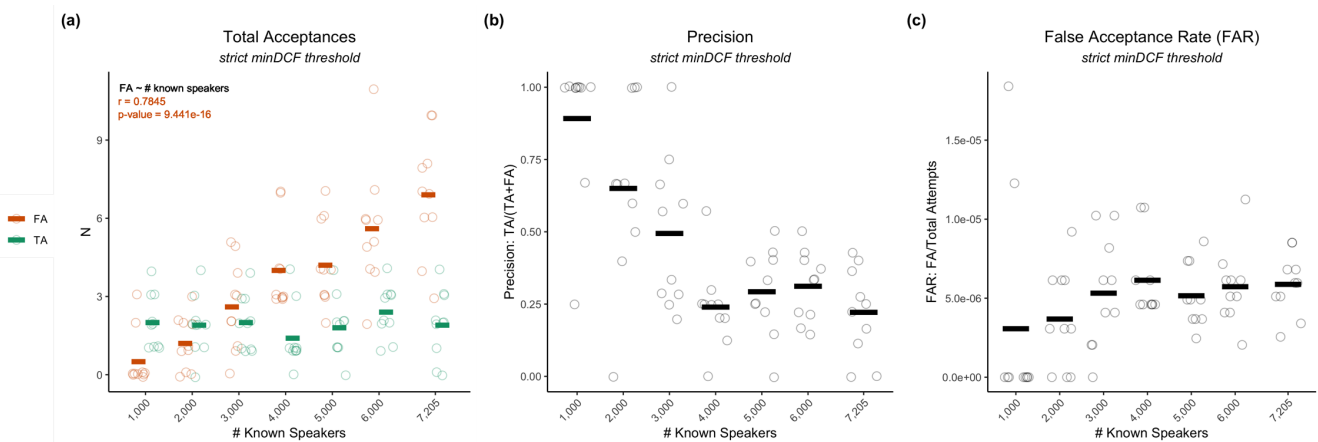

*Figure 6. Results for known speaker set with fixed speakers in unknown set using VoxCeleb. (a)* shows the breakdown of true and false acceptances when we change the number of known speakers and keep the unknown speaker set fixed, except for overlapping speakers. It also shows the Pearson's correlation coefficient and corresponding significance between false acceptances and number of known speakers. *(b)* shows the corresponding precision and *(c)* shows the corresponding false acceptance rates. Each run is plotted as a single circle, with red horizontal lines indicating the mean number of false acceptances, green horizontal lines indicating the mean number of true acceptances, and black horizontal lines indicating the mean precision or false acceptance rate.